\documentclass[prd,preprint,nofootinbib]{revtex4-1}

\usepackage{amssymb} 
\usepackage{amsmath}
\usepackage{xfrac}
\usepackage{bm}
\usepackage{rotating}
\usepackage{subcaption}
\usepackage{caption}
\usepackage{graphicx}
\renewcommand{\L}{\mathcal{L}}
\newcommand{\G}{\mathcal{G}}
\newcommand{\hf}{\frac{1}{2}}
\newcommand{\fb}[2]{\left(\frac{#1}{#2}\right)}
\newcommand{\abs}[1]{\left|#1\right|}

\begin{document}

\begin{flushright}
MAN/HEP/2018/06
\\
December 2018
\end{flushright}

\title{\Large{A Finite Measure for the Initial Conditions of Inflation}}

\author{Kieran Finn$^1$}\email{kieran.finn@manchester.ac.uk}\author{Sotirios Karamitsos$^{1,2}$}\email{s.karamitsos@lancaster.ac.uk}

\affiliation{$^1$School of Physics and Astronomy, University of Manchester, Manchester
 M13 9PL, United Kingdom\\
$^2$Consortium for Fundamental Physics, Physics Department, Lancaster University, Lancaster LA1 4YB, United Kingdom}




\begin{abstract}
We investigate whether inflation requires finely tuned initial conditions in order to explain the degree of flatness and homogeneity observed in the Universe. We achieve this by using the Eisenhart lift, which can be used to write any scalar field theory in a purely geometric manner. Using this formalism, we construct a manifold whose points represent all possible initial conditions for an inflationary theory. After equipping this manifold with a natural metric, we show that the total volume of this manifold is finite for a wide class of inflationary potentials. Hence, we identify a natural measure that enables us to distinguish between generic and finely tuned sets of initial conditions without the need for a regulator, in contrast to previous work in the literature. Using this measure, we find that the initial conditions that allow for sufficient inflation are indeed finely tuned. The degree of fine-tuning also depends crucially on the value of the cosmological constant at the time of inflation. Examining the example potential $V=\lambda\varphi^4+\Lambda$, we find that we require percent-level fine tuning if we allow the cosmological constant during inflation to be much larger than it is today. However, if we fix the cosmological constant to its presently observed value, the degree of fine tuning required is of order $10^{-29}$.
\end{abstract}

\maketitle

\section{Introduction}
\label{sec:intro}

The Hot Big Bang (HBB) model is one of the most successful in the field of cosmology. It explains many of the features of the Universe that we observe today, including the Hubble--Lema\^itre expansion law~\cite{Hubble:1929ig}, the cosmic microwave background (CMB) \cite{Penzias:1965wn}, as well as nucleosynthesis~\cite{Yang:1983gn} and large-scale structure formation \cite{Press:1973iz}. Nonetheless, the HBB model is not without its problems. Amongst other issues, it requires extremely finely tuned initial conditions in order to explain the large degree of homogeneity and flatness observed today. These problems, known as the horizon and flatness problems respectively, plagued cosmology until the formulation of the theory of inflation \cite{Guth:1980zm,Linde:1981mu,Albrecht:1982wi}. Inflation purports to solve these classic puzzles by a period of accelerated expansion that generically homogenises and flattens the observable Universe.

It is worth emphasising that the horizon and flatness problems are fundamentally problems of fine tuning. There is nothing in the HBB model to prevent a homogeneous and flat Universe, but this would require very specific initial conditions. Therefore, inflation can only be a real solution to the flatness and horizon problems if it requires less fine tuning than simply setting these initial conditions ``by hand''. Several arguments have been put forward to suggest that inflation does not require finely tuned initial conditions but instead happens generically. These arguments generally rest on two distinct principles: the volume weighting of different patches  and the universality of the inflationary attractor.

The volume weighting argument follows from the fact that patches of the Universe that inflate grow to be exponentially larger than those that do not \cite{Guth:2013sya}. Therefore, one might argue that even if only a tiny fraction of patches inflate, eventually \emph{most} of the Universe will be dominated by an inflationary patch. However, this argument breaks down when we consider a universe with infinite volume.
In this case both the inflationary and non-inflationary patches of the Universe are generically both infinite in volume \cite{PhysRevD.27.2848,Guth:2007ng,Brandenberger:2016uzh}, making their ratio ambiguous. In addition, this type of argument is plagued by the so-called \emph{youngness paradox}~\cite{Linde:1994gy,Guth:2007ng}. In an eternally expanding Universe, patches that stop inflating later than our patch by even one second outnumber it by a factor of $e^{10^{37}}$. It is therefore difficult to understand why our patch is so old in this paradigm.

An alternative argument for why inflation should be generic stems from its attractor nature \cite{Albrecht:1986pi,Brandenberger:1990xu}. It has been shown that for a scalar field in a flat Friedman--Lema\^itre--Robertson--Walker (FLRW) background, the slow-roll inflationary solution is an attractor in phase space~\cite{Salopek:1990jq,Liddle:1994dx}. One might therefore argue that no matter what region of phase space the Universe begins in, it will eventually  reach the slow-roll solution and inflation will ensue. However, even in the presence of the attractor, there are still some regions of the phase space that do not lead to observationally acceptable inflation. For example, the Universe may reach the attractor too late for sufficient inflation to occur ($N\gtrsim 60$ \mbox{e-foldings}). In addition, for curved universes, the attractor is not universal, and so there are some regions of phase space where inflation never occurs at all. In order to argue that inflation is still generic, one must argue that these regions are small. However, the space of initial conditions is infinite in extent. It is therefore unclear how we should define a ``small'' region. This difficulty is known as the \emph{measure problem}~\cite{Linde:1993nz,Linde:1993xx,GarciaBellido:1993wn, Linde:1995uf, Garriga:2001ri, Page:2008zh,Corichi:2010zp}.

There have been many attempts to quantify the fraction of phase space that leads to sufficient inflation for various models
\cite{Gibbons:1986xk,Chmielowski:1988zb, Coule:1994gd,Cho:1994vy,Stoeger:2004sn,Garriga:2005av,
Page:2006er,Gibbons:2006pa,Aguirre:2006na,Aguirre:2006ak,Clifton:2007bn,
Winitzki:2008yb}.
Most of these attempts implicitly rely on some method of regularising the infinite phase space that naturally arises when considering the totality of all possible initial conditions.\footnote{For an alternative method where a specific finite distribution on the fully infinite space is considered instead, see~\cite{Berezhiani:2015ola}.} For example, Gibbons, Hawking and Stewart (GHS)~\cite{Gibbons:1986xk} consider a cutoff in the Hubble parameter $H=\dot{a}/{a}$, showing that the resulting fraction is well-behaved in the limit where this cutoff is sent to infinity. Meanwhile, Gibbons and Turok (GT) \cite{Gibbons:2006pa} choose to take a cutoff in the curvature $k/a^2$ instead, again showing that their result is insensitive to the choice of cutoff value. However, as pointed out by Hawking and Page \cite{Hawking:1987bi}, the fraction of the phase space that leads to inflation is fundamentally ambiguous, as expected when taking the ratio of infinite quantities. As a result, it is possible to arrive at any desired limit by choosing different regularisation schemes. Indeed, the two sets of authors have contradictory conclusions: GHS find that inflation is almost guaranteed, while GT find inflation to be exponentially unlikely.

Clearly, a new outlook on the problem is required. In this paper, we will examine the Eisenhart lift \cite{eisenhart,Duval:1984cj,Cariglia:2015bla,Finn:2018cfs} as a possible approach to the measure problem. The Eisenhart-lift formalism allows any scalar field theory to be transformed (``lifted'') into a purely geometric system without altering its classical dynamics. Applying this formalism to Einstein gravity with a minimally coupled scalar field, we may construct a manifold whose geodesics exactly recreate all possible trajectories of the Universe, inflationary or otherwise. The problem of choosing initial conditions in the phase space is then reduced to that of choosing a point on the tangent bundle of this manifold. 
The tangent bundle comes with a natural metric~\cite{sasakimetric}, which can be used to define a natural measure on the space of initial conditions. This measure can in turn be used to rigorously define the volume of different regions of phase space. As we shall see, for a wide class of inflationary potentials, the total volume of the tangent bundle is finite. Thus, in this approach, the fraction of phase space that leads to sufficient inflation is well-defined from the start without the need to resort to regularisation.

This paper is laid out as follows: we begin in Section~\ref{sec:eisenhart} with a brief review of the Eisenhart-lift formalism, which may be used to construct a field-space manifold whose geodesics completely recover the classical dynamics of any homogeneous scalar field theory. In Section~\ref{sec:phase space metric}, we extend this idea by constructing a manifold for the full phase space of the theory. We use the natural metric on the phase space in Section~\ref{sec:measure} in order to place a measure on the allowed initial conditions. In Section~\ref{sec:inflation}, we calculate this measure for the case of a single scalar field in an FLRW background, and we show that the total measure is generically finite for a large class of inflationary potentials. We give a concrete example in Section~\ref{sec:example} by explicitly calculating the measure for a potential of the form $V(\varphi)=\lambda\varphi^4+\Lambda$ and quantifying the fraction of phase space that leads to sufficient inflation in this case. Finally, we present our conclusions and discuss possible directions for further research in Section~\ref{sec:discussion}.

Throughout this paper, we work in Planck units, for which $\hbar=c=M_P=1$.

\section{The Eisenhart Lift}
\label{sec:eisenhart}
The Eisenhart lift \cite{eisenhart,Duval:1984cj,Cariglia:2015bla} is a formalism in classical mechanics that may be used to write any system subject to a conservative force as an equivalent free system moving on a higher-dimensional curved manifold. This technique has been recently extended to scalar field theories \cite{Finn:2018cfs},  and can therefore be readily applied to the theory of inflation.

We begin our review of the Eisenhart lift by considering a field theory of $n$ homogeneous\footnote{The Eisenhart lift can also be applied to non-homogeneous field theories \cite{Finn:2018cfs}. However, this is unnecessary for the purpose of this paper, as we focus on the inflationary background.}
scalar fields $\varphi^i(t)$ (collectively denoted by~$\bm{\varphi}$), equipped with an arbitrary quadratic kinetic term and potential $V(\bm{\varphi})$. Such a theory is described by the Lagrangian
\begin{equation}\label{eq:force L}
{\cal L}\ =\ \hf k_{ij}(\bm{\varphi})\;\dot{\varphi}^i\dot{\varphi}^j\: -\: V(\bm{\varphi})\;,
\end{equation}
where the overdot denotes differentiation with respect to time $t$, the indices $i$ and $j$ run from~$1$ to $n$, and we use the Einstein summation notation, as we do throughout the rest of this paper. The equations of motion (EoMs) for this system may be found by varying~\eqref{eq:force L} with respect to the fields $\varphi^i$, and are
\begin{equation}\label{eq:force EoM}
\ddot{\varphi}^i\: +\: \Gamma^i_{jk}\,\dot{\varphi}^j\dot{\varphi}^k\ =\ -\, k^{ij}V_{,j}\;,
\end{equation}
where $,i$ indicates a derivative with respect to the field $\varphi^i$, $k^{ij}$ is the inverse of $k_{ij}$ (satisfying $k^{il}k_{lj}=\delta^i_j$), and we have defined 
\begin{equation}
\Gamma^i_{jk}\ =\ \hf k^{il}\,\Big(k_{jl,k}+k_{kl,j}-k_{jk,l}\Big).
\end{equation}
We note that~\eqref{eq:force EoM} bears a striking resemblance to a geodesic equation with an additional force term. Indeed, we may easily describe the evolution of the system in the language of differential geometry. The fields $\varphi^i$ now take on the role of coordinates in some particular chart of a (possibly curved) manifold known as the \emph{field space} \cite{GrootNibbelink:2000vx}. This manifold comes equipped with a metric $k_{ij}$, which can be used to construct the connection through the Christoffel symbols $\Gamma^i_{jk}$. In this interpretation, the system at a given moment in time is described by a point on this manifold and its evolution is described by a trajectory.

The Eisenhart-lift formalism allows us to extend this geometric interpretation by constructing a higher-dimensional field space manifold for which the trajectories,~\eqref{eq:force EoM}, are geodesics and there is no need for an external force field. To this end, we introduce a new \emph{fictitious} field $\chi$ to our theory and modify the Lagrangian as follows:
\begin{equation}\label{eq:new lagrangian}
{\cal L}\ =\ \hf k_{ij}(\bm{\varphi})\;\dot{\varphi}^i\dot{\varphi}^j\: +\: \hf \frac{M^4}{V(\bm{\varphi})}\,\dot{\chi}^2\;,
\end{equation}
where $M$ is an arbitrary mass scale, introduced to keep the dimensions of the fields consistent.  This ``lifted'' Lagrangian \eqref{eq:new lagrangian} can be written concisely as
\begin{equation}\label{eq:lifted lagrangian}
\L=\hf\,G_{AB}\dot{\phi}^A\dot{\phi}^B,
\end{equation}
where $\phi^A=\{\varphi^i,\chi\}$ (collectively denoted by $\bm \phi$), the indices $A$ and $B$ take on the values $1\leq A,B\leq n+1$, and $G_{AB}$ is given by
\begin{equation}
G_{AB}\ \equiv\ 
\begin{pmatrix}  
k_{ij}	&	0\\
0		&	\dfrac{M^4}{V}
\end{pmatrix}.
\end{equation}
Crucially, the Lagrangian~\eqref{eq:lifted lagrangian} is identical to that of a free particle moving on a curved manifold equipped with metric $G_{AB}$. Thus, the trajectories of the system will be exactly described by geodesics of this manifold. 

We now show that the geodesics for the manifold corresponding to the lifted system reduce to the EoMs for the original system~\eqref{eq:force EoM}. We may find the geodesics by varying~\eqref{eq:new lagrangian} with respect to $\varphi^i$ and $\chi$, which yields the following EoMs:
\begin{align}
\ddot{\varphi}^i\: +\: \Gamma^i_{jk}\,\dot{\varphi}^j\dot{\varphi}^k\ &=\ -\, \hf k^{ij}V_{,j}\,\frac{M^4\dot{\chi}^2}{V^2},\label{eq:varphi eom}
\\
\frac{d}{dt}\fb{\dot{\chi}}{V(\bm{\varphi})}\ &=\ 0\;.\label{eq:chi eom}
\end{align}
Equation~\eqref{eq:chi eom} has the following class of solutions:
\begin{equation}\label{eq:chidot sol}
\dot{\chi}\ =\ A \frac{V(\bm{\varphi})}{M^2}\ ,
\end{equation}
parametrised by a real constant $A$. When the fictitious field $\chi$ satisfies~\eqref{eq:chidot sol}, then the EoM~\eqref{eq:varphi eom} becomes
\begin{equation}
\ddot{\varphi}^i\: +\: \Gamma^i_{jk}\dot{\varphi}^j\dot{\varphi}^k\ =\ 
-\,\frac{A^2}{2} k^{ij}V_{,j}\; .\label{eq:varphi A eom}
\end{equation}
We note that the EoMs~\eqref{eq:varphi A eom} (which arise purely as a consequence of the geometric structure of the field space) are identical to the original EoMs~\eqref{eq:force EoM}, provided that~$A$ satisfies the \emph{Eisenhart condition}:
\begin{equation}\label{eq:eisenhart condition}
A^2=\fb{\dot{\chi}M^2}{V(\bm{\varphi})}^2=2.
\end{equation}
Thus we have achieved the desired result; classically, the two systems described by the two Lagrangians~\eqref{eq:force L} and~\eqref{eq:new lagrangian} are identical.

The Eisenhart condition~\eqref{eq:eisenhart condition} should not be thought of as a condition on the fields or the initial conditions. In fact, even if the Eisenhart condition is not satisfied, the trajectory of the lifted system through the field space is still the same. The only change is that the system will evolve faster ($A^2>2$) or slower ($A^2<2$) than it would under the original EoMs~\eqref{eq:force EoM}. However, the classical dynamics of the lifted theory are invariant under affine time reparametrisations. Therefore, we can always scale time in such a way that~\eqref{eq:eisenhart condition} is satisfied. On the contrary, the system described by the original Lagrangian~\eqref{eq:force L} is not symmetric under time rescalings, as we can see from its EoMs~\eqref{eq:force EoM}. This means that the EoMs of the two systems can only match for a specific parametrisation. This parametrisation is the one in which the Eisenhart condition is satisfied.

\section{The Phase Space Manifold}
\label{sec:phase space metric}

In the previous section, we demonstrated that any homogeneous scalar field theory (such as inflation) can be written in terms of geodesic motion on a lifted field space manifold. We should be able to use the volume of this manifold as a measure to distinguish between generic sets of initial conditions from finely tuned ones. However, we must take into account that the EoMs for the system are of second order. Therefore, the initial conditions include not only the initial value of the fields, but also the initial values of their time derivatives. Thus, the initial conditions will live in the \emph{phase space} manifold. The topic of this section is the geometric structure of this manifold.

The phase space manifold is the tangent bundle of the field space. The natural metric on a tangent bundle is the Sasaki metric \cite{sasakimetric}. This is the unique metric that satisfies the following three properties:
\begin{enumerate}
\item It reduces to the metric of the original manifold (in our case, the field-space manifold) when restricted to the origin of the tangent space.
\item It leaves all tangent spaces flat.
\item The line element it defines is reparametrisation invariant.
\end{enumerate}

It is important to note that the quantity $d\dot{\phi}^A$ is not a field-space vector and thus the quantity $G_{AB}d\dot{\phi}^Ad\dot{\phi}^B$ (which we may have naively used to build our line element) is not reparametrisation invariant. Instead, we employ the field-space covariant variation
\begin{equation}
D\dot{\phi}^A=d\dot{\phi}^A+\Gamma^A_{BC}\dot{\phi}^B d\phi^C,
\end{equation}
which is now a fully covariant field space vector. Here, $\Gamma^A_{BC}$ are the Christoffel symbols for the field-space metric $G_{AB}$. Using this variation, we may write the Sasaki line element for the phase space as\footnote{Note that, by dimensionality, there must be a mass scale multiplying the second term in~\eqref{eq:phase space line element}, which we have chosen to set to the Planck mass. This choice is arbitrary, since any other choice can be rescaled by simply dilating our time coordinate appropriately. Since time dilation is a symmetry of the system, this choice does not affect our results.}
\begin{equation}\label{eq:phase space line element}
ds^2=G_{AB}d\phi^Ad\phi^B+G_{AB}D\dot{\phi}^AD\dot{\phi}^B.
\end{equation}
The explicit form of the phase space metric in the ${\Phi^\alpha=\{\bm\phi,\dot{\bm\phi}\}}$ coordinate basis is
\begin{equation}\label{eq:phase space metric}
\G_{\alpha\beta}=\left(\begin{array}{cc}
G_{AB}+G_{CD}\Gamma^C_{AE}\Gamma^D_{BF}\dot{\phi}^E\dot{\phi^F}&G_{CB}\Gamma^C_{AD}\dot{\phi}^D\\
G_{AC}\Gamma^C_{DB}\dot{\phi}^D&G_{AB}
\end{array}\right),
\end{equation}
where $\alpha,\beta$ run from 1 to $2n$.

It should be clear that the metric given in~\eqref{eq:phase space metric} satisfies all the above requirements. First, it reduces to $G_{AB}$ at the origin of the tangent space where $\dot{\bm\phi}=0$. Second, since $G_{AB}$ does not depend on derivatives $\dot{\bm\phi}$, the Riemann curvature tensor will vanish for the tangent space of any point in the field space. Thus, all tangent spaces are flat, as prescribed. Finally, by construction, the line element in~\eqref{eq:phase space line element} is reparametrisation invariant, since it is constructed from the inner product of field-space vectors.

\section{A Measure on Initial Conditions}
\label{sec:measure}

In the previous section, we arrived at a natural metric for the phase space manifold. We are now equipped to use this metric in order to write a measure, which we may use to assign weights to different initial conditions. The phase space metric~\eqref{eq:phase space metric} can be used to explicitly write the invariant volume element for the phase space as
\begin{equation}\label{eq:phase space measure}
d\Omega=\sqrt{\det(\G_{\alpha\beta})} \, d^{2n}\Phi \; =\; \det(G_{AB})d^n \,\dot{\phi}\,d^n\phi.
\end{equation}
Note that we get an additional factor of $\sqrt{\det(G_{AB})}$ compared to what we might expect from the field space alone. We propose to use this volume element as a measure on the phase space.

At this stage, we should compare the measure~\eqref{eq:phase space measure} to another commonly used measure in the literature, the Liouville measure \cite{Gibbons:1986xk,Gibbons:2006pa}, which is often used because it is invariant under time evolution. In order to construct the Liouville measure, we may take advantage of the symplectic nature of the phase space. Any symplectic manifold by definition comes equipped with a symplectic form, which may be written as
\begin{equation}
\omega=dp_A\wedge dq^A.
\end{equation}
Here, $\wedge$ denotes the wedge product and $q^i$ and $p_i$ are the generalised coordinates and canonical momenta respectively. The Liouville volume element is then
\begin{equation}\label{eq:Liouville measure}
d\Omega_L = \frac{(-1)^{n(n-1)/2}}{n!}\omega^{\wedge n} =d^n q \, d^n p,
\end{equation}
where $\omega^{\wedge n}$ denotes the wedge product being taken $n$ times.

Despite the fact that the Liouville measure given in~\eqref{eq:Liouville measure} and the lifted measure  given in~\eqref{eq:phase space measure} appear to be different, they are in fact identical for the system described by the Lagrangian~\eqref{eq:lifted lagrangian}. For this system, the canonical variables $q^A$ and $p_A \equiv \partial \mathcal{L}/\partial \dot q^A$ are
\begin{equation}
q^A=\phi^A,\hspace{4em}
p_A=G_{AB}\dot{\phi}^B.
\end{equation}
The symplectic form~\eqref{eq:Liouville measure} is thus
\begin{equation}\label{eq:two form}
\omega=G_{AB} d\dot{\phi}^A\wedge d\phi^B+G_{AB,C}\, \dot{\phi}^Ad\phi^C\wedge d\phi^B.
\end{equation}
When using~\eqref{eq:Liouville measure} to construct the Liouville measure, it is clear that the second term will not contribute, since all terms that include it must involve a wedge product of at least $n+1$ $d\phi$ terms. Such terms will necessarily, by the pigeonhole principle, include the wedge product of a quantity with itself, and will thus vanish. The only non-vanishing term is then
\begin{equation}\label{eq:liouville expanded}
d\Omega_L=G_{1A}G_{2B}...G_{nZ}d\dot{\phi}^A\wedge d\dot{\phi}^B...\wedge d\dot{\phi}^Z\wedge d\phi^1\wedge d\phi^2...\wedge d\phi^n.
\end{equation}
Using the following identity:
\begin{equation}
d\dot{\phi}^A\wedge d\dot{\phi}^B...\wedge d\dot{\phi}^Z=\epsilon^{AB...Z}d\dot{\phi}^1\wedge d\dot{\phi}^2...\wedge d\dot{\phi}^n,
\end{equation}
we can rewrite~\eqref{eq:liouville expanded} as
\begin{equation}
d\Omega_L=\epsilon^{AB...Z}\, G_{1A}G_{2B}...G_{nZ} \,d^n\phi d^n\dot{\phi}=\det G \, d^n  {\phi} \, d^n \dot \phi = d\Omega.
\end{equation}
We have therefore shown that the two measures are equivalent.

Unfortunately, we cannot immediately use the measure~\eqref{eq:phase space measure} over the entire phase space to weight the initial conditions for a theory described by a diffeomorphism-invariant Lagrangian with an Einstein-Hilbert term (such as the theory of inflation). In such a theory, the variation of the action with respect to the $00$ component of the spacetime metric $g_{\mu\nu}$ (often known as the lapse, $N_L$) yields the \emph{Hamiltonian constraint}. This is an algebraic, not dynamical, constraint and must therefore be satisfied by all physical configurations of the fields, which of course includes the initial conditions. Thus, the only physically relevant part of the phase space manifold is the $2n-1$ dimensional hypersurface on which the Hamiltonian constraint is satisfied. We call this the \emph{Hamiltonian hypersurface}.

The Hamiltonian hypersurface has a metric induced on it by virtue of being embedded in the phase-space manifold. This induced metric is given by
\begin{equation}\label{eq:constrained metric}
\widetilde{\G}_{ab}=\frac{\partial F^\alpha}{\partial {\widetilde \Phi}^a}\frac{\partial  F^\beta}{\partial {\widetilde \Phi}^b}\G_{\alpha\beta},
\end{equation}
where $\widetilde{\Phi}^a$ with $1\leq a,b\leq 2n-1$ (collectively $\widetilde{\bm\Phi}$) are coordinates on the Hamiltonian hypersurface. The functions $F^\alpha$ encode the embedding through
\begin{align}
\Phi^\alpha =F^\alpha(\widetilde{\bm \Phi}),
\end{align}
where, as before, $\Phi^\alpha=\{\bm\phi,\dot{\bm\phi}\}$. With the above definition of the induced metric, it is then possible to write the volume element of the Hamiltonian hypersurface as
\begin{equation}\label{eq:constrained measure}
d\widetilde{\Omega}=\sqrt{\det(\widetilde{\G}_{ab})}\, d^{2n-1}\widetilde{\bm\Phi}.
\end{equation}

The volume element~\eqref{eq:constrained measure} can be used as a measure on the set of all possible initial conditions for the scalar field theory described by the Lagrangian~\eqref{eq:lifted lagrangian}. In this paper, we postulate that this measure can also be used for the classically equivalent system described by Lagrangian~\eqref{eq:force L}. In the following section, we will use it to distinguish between generic and finely tuned sets of initial conditions in such a theory.

\section{Application to Inflation}
\label{sec:inflation}
We are now ready to apply the technology developed in the previous sections to the theory of inflation. We consider a universe described by Einstein gravity and dominated by a single minimally-coupled scalar field $\varphi$. We thus take the action to be
\begin{equation}
S = \int d^4 x\sqrt{-g} \left( -\frac{R}{2} + \frac{1}{2} (\partial_\mu \varphi)(\partial^\mu \varphi) - V(\varphi)\right),
\end{equation}
where $R$ is the Ricci scalar, $V(\varphi)$ is the inflationary potential, and Lorentz indices are raised and lowered with the help of the spacetime metric $g_{\mu\nu}$ with determinant $g$.

In this paper, we shall work in the minisuperspace approximation, which posits that the Universe is described by an FLRW Universe, with metric
\begin{equation}\label{eq:FRW metric}
ds^2=N_L^2(\tau) d\tau^2-a^2(\tau)\left(\frac{1}{1-kr^2}dr^2+r^2d\theta^2+r^2\sin^2\theta \, d\phi^2\right),
\end{equation}
where $\tau$ is the time coordinate in this chart, $N_L(\tau)$ is the lapse function, $a(\tau)$ is the scale factor, and $k$ is the extrinsic curvature of the spatial part of the metric. Moreover, we assume the inflaton to be homogeneous
(a valid approximation for a small patch). Of course, the reader may accuse us of cheating somewhat by assuming homogeneity from the outset, since homogeneity is precisely what we wish to achieve through inflation. However, inflation requires only a small patch of the Universe to be homogeneous in order to \hbox{begin~\cite{Barrow:1984zz,Jensen:1986nf,Goldwirth:1989pr,Bruni:2001pc}}.  Therefore, this assumption is much weaker than the one made in the HBB model. Estimating the
likelihood 
of such a patch arising is an interesting question in its own right, but is beyond the scope of this paper.\footnote{For discussions of this topic and how inflation fares in an inhomogeneous Universe, see~\cite{Berezhiani:2015ola,East:2015ggf,Kleban:2016sqm,Clough:2016ymm,
Clough:2017efm}.}
We note that adding inhomogeneities makes inflation harder to achieve, since we must produce a homogeneous patch in addition to sufficient inflation of that patch.

In addition, we make the simplifying assumption that $k = 0$. Although this assumption restricts the range of validity of our calculations, we note that the inflationary attractor is most effective for flat universes. Indeed, with $k=0$, all trajectories approach the attractor as $t\to\infty$, something which is not guaranteed for $k\neq0$. As such, we expect that this approximation will further favour inflation. Therefore, the results of the following sections should be thought of as upper bounds on the fraction of phase space that allows inflation.

Under these assumptions, the Lagrangian for this system is
\begin{equation}
\label{eq:non-lifted L}
L=-3\frac{a}{N_L}\fb{da}{d\tau}^2+\frac{a^3}{2N_L}\fb{d\varphi}{d\tau}^2-N_L a^3 V(\varphi).
\end{equation}
We can now apply the Eisenhart lift described in Section~\ref{sec:eisenhart} in order to obtain a purely kinetic Lagrangian that will yield the same classical dynamics as~\eqref{eq:non-lifted L}. The lifted Lagrangian is
\begin{equation}
\label{eq:lifted L with N}
L=\frac{1}{N_L}\left[-3a\fb{da}{d\tau}^2+\hf a^3\fb{d\varphi}{d\tau}^2+\hf\frac{1}{a^3V(\varphi)}\fb{d\chi}{d\tau}^2\right],
\end{equation}
where we have chosen the arbitrary mass scale labeled $M$ in Section~\ref{sec:eisenhart} to be $M=1$ in Planck units. No observable quantities can depend on this choice, since scaling $M$ simply amounts to rescaling the fictitious field $\chi$.

We can simplify~\eqref{eq:lifted L with N} by changing coordinates such that $dt= \sqrt{N_L}d\tau$. In these coordinates, the lifted Lagrangian becomes 
\begin{equation}\label{eq:lifted L}
L=-3a\dot{a}^2+\hf a^3\dot{\varphi}^2+\hf\frac{1}{a^3V(\varphi)}\dot{\chi}^2,
\end{equation}
where as before, the overdot denotes differentiation with respect to $t$.
Even though we have eliminated the dependence on $N_L$, we must remember that varying the action with respect to it leads to the following algebraic equation of motion:
\begin{equation}\label{eq:Hamiltonian constraint}
{\cal H}=-3a\dot{a}^2+\hf a^3\dot{\varphi}^2+\hf\frac{1}{a^3V(\varphi)}\dot{\chi}^2=0,
\end{equation}
where ${\cal H}$ is the Hamiltonian of the system. This is the Hamiltonian constraint referred to in Section~\ref{sec:measure}. Only initial conditions that satisfy~\eqref{eq:Hamiltonian constraint} are physically allowed.

As explained in Section~\ref{sec:eisenhart}, the evolution of this system can be described as a geodesic in a three-dimensional field space parametrised by coordinates $\phi^A=\{a,\varphi,\chi\}$, and equipped with a field-space metric given by
\begin{equation}
G_{AB}=\begin{pmatrix}  
-6\,a	&	0&0\\
0		&	a^3&0\\
0&0&\dfrac{1}{a^3V(\varphi)}
\end{pmatrix}.
\end{equation}
Let us derive explicit forms for these geodesics. The equation of motion for the fictitious field $\chi$ is
\begin{equation}
\frac{d}{dt}\fb{\dot{\chi}}{a^3V(\varphi)}=0.
\end{equation}
We thus find that the field $\chi$ satisfies
\begin{equation}\label{eq:infl eisenhart condition}
\frac{\dot{\chi}}{a^3V(\varphi)}=A, 
\end{equation}
where $A$ is a constant. As described in Section~\ref{sec:eisenhart}, we require that $A^2=2$ in order to satisfy the Eisenhart condition~\eqref{eq:eisenhart condition}. This is crucial in order for the rest of the EoMs to reproduce those arising from~\eqref{eq:non-lifted L}. As mentioned earlier, we can always satisfy this condition by choosing an affine parametrisation for time.

With the Eisenhart condition satisfied, the EoMs for $\varphi$ and $a$ become 
\begin{align}
\label{eq:friedmann phi}
 \ddot{\varphi} \, + \, 3H\dot{\varphi}& \, + \,V'(\varphi)=0,
\\
\label{eq:friedmann a}
H^2 \,+ \, 2\frac{\ddot{a}}{a} & \,= \, \hf\dot{\varphi}^2 \,+ \,V(\varphi),
\end{align}
where $H\equiv \frac{\dot{a}}{a}$ is the Hubble parameter and the Hamiltonian constraint~\eqref{eq:Hamiltonian constraint} becomes
\begin{equation}\label{eq:unlifted Hamiltonian constraint}
-3H^2+\hf \dot{\varphi}^2+V(\varphi)=0.
\end{equation}
As expected, these are simply the Friedmann equations.

Following the process outlined in Section~\ref{sec:phase space metric}, we may now construct a six-dimensional phase space manifold with coordinates ${\Phi^\alpha=\{a,\varphi,\chi,\dot{a},\dot{\varphi},\dot{\chi}\}}$. The metric of this phase space can be calculated from~\eqref{eq:phase space metric}, and is shown explicitly in Appendix~\ref{sec:app phase space metric}. We note that each point on this manifold corresponds to a different possible initial configuration of the Universe, but only points satisfying the Hamiltonian constraint~\eqref{eq:Hamiltonian constraint} correspond to physically allowed initial conditions. However, there is a redundancy in our description of the system due to its symmetries, which we now turn our attention to.

The first symmetry, which is present in both the original and the lifted system, is a spatial dilation symmetry. We note that for a flat homogeneous universe, there is no characteristic scale. Thus, the transformation $r\to c r$ must correspond to a symmetry. In terms of the coordinates of phase space, this is equivalent to the following transformations:
\begin{equation}\label{eq:spatial dilation}
a\to ca,\hspace{1em}\chi\to c^3\chi,\hspace{1em}\dot{a}\to c\dot{a},\hspace{1em}\dot{\chi}\to c^3\dot{\chi},
\end{equation}
where the $\chi$ transformation follows from the Eisenhart condition~\eqref{eq:infl eisenhart condition}. Notice that the Lagrangian~\eqref{eq:lifted L} is not left unchanged by these transformations -- it transforms as $L\to c^3L$. However, any constant factor in front of $L$ will drop out of the Euler--Lagrange equations. Therefore, the EoMs will be unaffected by this transformation \cite{Sloan:2018lim}.

The lifted system~\eqref{eq:lifted L} features two additional symmetries. There is a shift symmetry in the fictitious field $\chi$. Indeed, the Lagrangian~\eqref{eq:lifted L} is invariant under the transformation
\begin{equation}\label{eq:shift symmetry}
\chi\to\chi+c.
\end{equation}
Furthermore, we have the time dilation symmetry which we briefly alluded to earlier. Scaling our time coordinate by a constant factor ($t\to c\,t$) will cause the coordinates in phase space to transform as
\begin{equation}\label{eq:time dilation}
\dot{a}\to\frac{1}{c}\dot{a},\hspace{1em}\dot{\varphi}\to\frac{1}{c}\dot{\varphi},\hspace{1em}\dot{\chi}\to\frac{1}{c}\dot{\chi}.
\end{equation}
These transformations cause the Langrangian~\eqref{eq:lifted L} to scale by $L\to \frac{1}{c^2}L$, leaving the EoMs invariant.

The time-dilation symmetry may be understood more intuitively as follows. In the Eisenhart-lifted theory, the classical evolution of the Universe is described by a geodesic on the field-space manifold. Speeding up or slowing down time will not alter this geodesic, but only change how fast the system evolves along it. However, an observer within a universe described by such a theory has no way to tell how fast a particular trajectory evolves. Any clocks or time-keeping devices will be constructed out of the fields, and will thus be sped up or slowed down in exactly the same way as everything else. 
Indeed, if everything in the Universe, including your watch, suddenly started moving twice as fast would you notice?

The above three symmetries tell us that there are redundancies in our choice of initial conditions. Only three combinations of the initial values $a_0$, $\varphi_0$, $\chi_0$, $\dot{a}_0$, $\dot{\varphi}_0$ and $\dot{\chi}_0$ are relevant to the resulting evolution of the Universe. The other three are irrelevant thanks to the symmetries described above and are akin to gauge parameters. We further note that of the three relevant combinations, one must be fixed by~\eqref{eq:Hamiltonian constraint}. Thus, we expect the physically distinct initial conditions to be parametrised by two parameters.

Having identified the symmetries of the system, we may change to a coordinate system that manifestly distinguishes between physically relevant and irrelevant degrees of freedom. We start by isolating the transformations~\eqref{eq:spatial dilation} and~\eqref{eq:shift symmetry} with the help of a system of coordinates $(\lambda, H, \widetilde \chi, H_\chi)$ defined as
\begin{align}
\begin{aligned}
\lambda &\equiv\ln a, 
&
\dot{\lambda}&=\frac{\dot{a}}{a}\equiv H,\\
\widetilde{\chi}&\equiv\frac{\chi}{a^3},
&
H_{\chi}&\equiv\frac{\dot{\chi}}{a^3}.
\end{aligned}
\end{align}
Of these coordinates, $\lambda$ is the only one affected by the transformation~\eqref{eq:spatial dilation} and $\widetilde{\chi}$ is the only one affected by the transformation~\eqref{eq:shift symmetry}. Thus, the initial values of these two parameters are irrelevant. Proceeding in a similar manner, we may also define the following coordinate system in order to isolate the effect of the transformation~\eqref{eq:time dilation}:
\begin{align}
H&\equiv \frac{\rho}{\sqrt{6}}\cos\alpha,\label{eq:polar H}\\
\dot{\varphi}&\equiv \rho\sin\alpha\cos\beta,\label{eq:polar phi}\\
H_{\chi}&\equiv \rho \sqrt{V}\sin\alpha\sin\beta.\label{eq:polar chi}
\end{align}
Now only the coordinate $\rho$ is affected by this transformation.

In the coordinate chart ${\Phi^\alpha=\{\lambda,\varphi,\widetilde{\chi},\rho,\alpha,\beta\}}$, the Hamiltonian is 
\begin{equation}
{\cal H}=-\hf e^{3\lambda}\rho^2\cos(2\alpha)
\end{equation}
and thus the Hamiltonian constraint~\eqref{eq:Hamiltonian constraint} is simply
\begin{equation}\label{eq:ham constraint alpha}
\alpha=\frac{\pi}{4}.
\end{equation}
We must restrict ourselves to the five-dimensional hypersurface on which~\eqref{eq:ham constraint alpha} is satisfied. This Hamiltonian hypersurface can be parametrised by ${\widetilde{\Phi}^a=\{\lambda,\varphi,\widetilde{\chi},\rho,\beta\}}$, and has an induced metric given by~\eqref{eq:constrained metric}. However, we find the metric on the Hamiltonian hypersurface to be singular with
\begin{equation}
\widetilde{\G}_{\alpha 4}=\widetilde{\G}_{4\alpha}=0\label{eq:singular metric}
\end{equation}
for all values of the index $\alpha$. Here $4$ corresponds to the $\rho$ coordinate, whose initial value is irrelevant to observables. As a consequence, the invariant volume element of this manifold vanishes. Therefore, we cannot use it to construct a measure to help us distinguish between finely tuned from generic sets of initial conditions.

To remedy the above problem, we use a regularization technique. We consider a hypersurface very close to the Hamiltonian hypersurface, on which the Hamiltonian is
\begin{equation}
{\cal H}=\epsilon
\end{equation}
and  take the limit as $\epsilon\to 0$. The induced metric on this modified Hamiltonian hypersurface is non-singular and has an invariant volume element given by
\begin{equation}
d\widetilde{\Omega}(\epsilon)
\;
=
\;
 \sqrt{\frac{6e^{15\lambda}\rho^2\epsilon}{e^{3\lambda}\rho^2-2\epsilon}}
 \;
 \frac{1}{\sqrt{V(\varphi)}}\; d\lambda\, d\varphi\, d\widetilde{\chi}\, d\rho\, d\beta.\label{eq:full measure}
\end{equation}
Notice that the total volume of the manifold is proportional to $\sqrt{\epsilon}$, and vanishes as $\epsilon$ tends to zero. However, as we shall see, this dependence on $\epsilon$ will drop out for any ratio of physically distinguishable sets of initial conditions. We emphasise again that only the initial values of $\varphi$ and $\beta$ are relevant to the evolution. The initial values of $\lambda$, $\widetilde{\chi}$ and $\rho$ have no observable effect and can be changed arbitrarily by the symmetry transformations~\eqref{eq:spatial dilation},~\eqref{eq:shift symmetry},  and~\eqref{eq:time dilation} respectively.

As described in Section~\ref{sec:measure}, we can distinguish generic sets of initial conditions from finely tuned ones by the volume they take up on this hypersurface. This amounts to using~\eqref{eq:full measure} as a measure on the space of initial conditions. 

The total volume of the Hamiltonian hypersurface is infinite, potentially ruining our ability to compare the size of regions in an unambiguous way. However, note that if $V(\varphi)$ approaches infinity faster than $\varphi^2$ as $\varphi\to\infty$ and there is a non-zero cosmological constant such that $V(\varphi)$ is strictly positive,\footnote{This is true for the simplest, still-viable inflationary potential $\lambda\varphi^4$ but may not be true for potentials with infinite plateaus. For such potentials, the measure may still require regulating, for instance by adding a wall at the end of the plateau.} the following integral is finite:
\begin{equation}
{\cal N}\equiv\int_{-\infty}^\infty \frac{1}{\sqrt{V(\varphi)}}\,d\varphi.
\end{equation}
In this case the manifold is infinite only in the directions of  $\lambda$, $\widetilde{\chi}$ and~$\rho$, which are precisely the three physically meaningless coordinates. We can thus integrate out these directions to recover the following well-defined distribution for the initial values of the physically relevant coordinates $\varphi$ and $\beta$ on the manifold:
\begin{widetext}
\begin{align}
P(\varphi_1,\varphi_2,\,\beta_1,\beta_2)&= \lim_{\epsilon\to0}\; \lim_{\Lambda_\lambda,\Lambda_{\widetilde{\chi}}\Lambda_\rho\to \infty}\frac{\int_{\widetilde{\chi}=-\Lambda_{\widetilde{\chi}}}^{\widetilde{\chi}=\Lambda_{\widetilde{\chi}}}\int_{\lambda=-\Lambda_\lambda}^{\lambda=\Lambda_\lambda}\int_{\rho=0}^{\rho=\Lambda_\rho}\int_{\varphi=\varphi_1}^{\varphi=\varphi_2}\int_{\beta=\beta_1}^{\beta=\beta_2}   d\widetilde{\Omega}(\epsilon)}{\int_{\widetilde{\chi}=-\Lambda_{\widetilde{\chi}}}^{\widetilde{\chi}=\Lambda_{\widetilde{\chi}}}\int_{\lambda=-\Lambda_\lambda}^{\lambda=\Lambda_\lambda}\int_{\rho=0}^{\rho=\Lambda_\rho}\int_{\varphi=-\infty}^{\varphi=\infty}\int_{\beta=0}^{\beta=2\pi}   d\widetilde{\Omega}(\epsilon)}\\
&= \frac{\beta_2-\beta_1}{2\pi {\cal N}}\int_{\varphi=\varphi_1}^{\varphi=\varphi_2}\frac{1}{\sqrt{V(\varphi)}} d\varphi,\label{eq:probability}
\end{align}
\end{widetext}
where $P(\varphi_1,\varphi_2,\,\beta_1,\beta_2)$ is the fraction of phase space for which ${\varphi_1<\varphi<\varphi_2}$ and ${\beta_1<\beta<\beta_2}$.

The initial value of $\beta$ can be translated back into the more conventional coordinates of $\dot{\varphi}$ and $H$ with the help of the Eisenhart condition~\eqref{eq:infl eisenhart condition}, the Hamiltonian constraint~\eqref{eq:ham constraint alpha} and the definitions~\eqref{eq:polar H} and~\eqref{eq:polar phi}. These combine to give
\begin{align}
H&= \frac{1}{\sin\beta}\sqrt{\frac{1}{3}V(\varphi)},
\\
\dot{\varphi}&= \frac{1}{\tan\beta}\sqrt{2V(\varphi)}.\label{eq:pdot of beta}
\end{align}
Notice that, by construction,~\eqref{eq:unlifted Hamiltonian constraint} is automatically satisfied with these relations.

\section{Example: $\varphi^4$ inflation}
\label{sec:example}

As a concrete example, we turn our attention to the common chaotic inflationary potential,
\begin{equation}\label{eq:phi_4}
V(\varphi)=\lambda\varphi^4+\Lambda.
\end{equation}
The cosmological constant (which we take to be strictly positive) is usually ignored during inflation, but here we find that it is required for the volume of the phase-space manifold to be finite. Although $\varphi^4$ inflation is disfavoured at the $2\sigma$ level by the non-observation of tensor modes in the CMB, it is still worth analysing as the simplest inflationary potential that is still viable. The value of the quartic coupling $\lambda$ is set by the amplitude of scalar perturbations in the CMB \cite{Akrami:2018odb} and has a value of $\lambda\approx5\times 10^{-14}$.

Note that we do not constrain the value of $\Lambda$. Even though today the cosmological constant is measured to be $\Lambda=2.846\times 10^{-122}$~\cite{Aghanim:2018eyx} in Planck units, we allow for the possibility that it was larger at early times in order to study its effect on the required degree of fine tuning. However, we cannot have an arbitrarily large cosmological constant. If $\Lambda$ is too large, then the slow roll parameter
\begin{equation}
\epsilon=\hf \fb{V'(\varphi)}{V(\varphi)}^2=\frac{8\lambda^2\varphi^6}{\lambda^2\varphi^8+2\lambda\Lambda\varphi^4+\Lambda^2}
\end{equation}
can never surpass $\epsilon=1$ and thus inflation can never end. This condition forces us to consider~$\Lambda<\frac{27}{4}\lambda$.

\begin{centering}
\begin{figure*}[!t]
\makebox[\textwidth][c]{
        \begin{subfigure}[h]{0.5\textwidth}               
                \includegraphics[width=\textwidth]{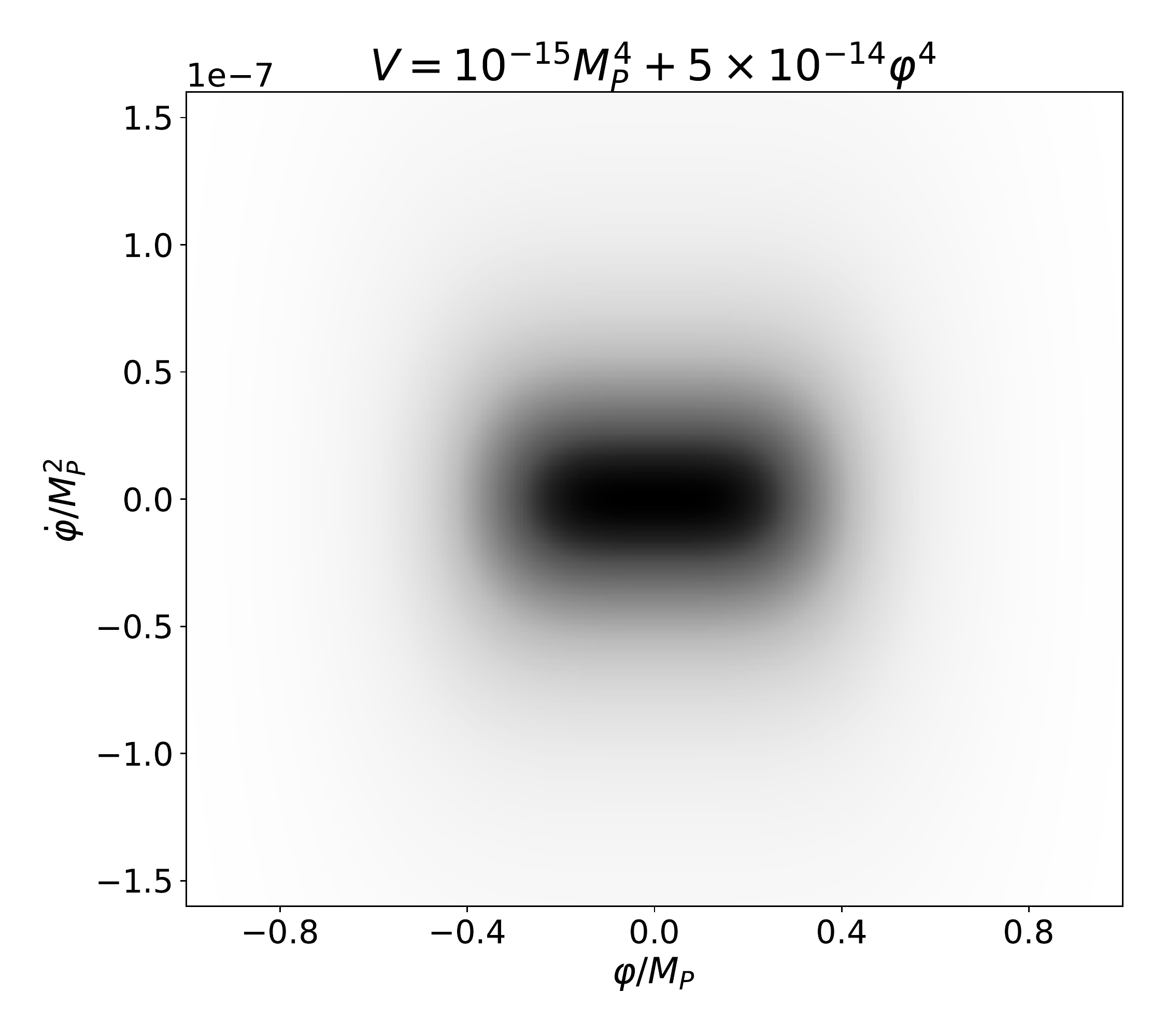}
        \end{subfigure} %
        ~ 
        \begin{subfigure}[h]{0.5\textwidth}
                \includegraphics[width=\textwidth]{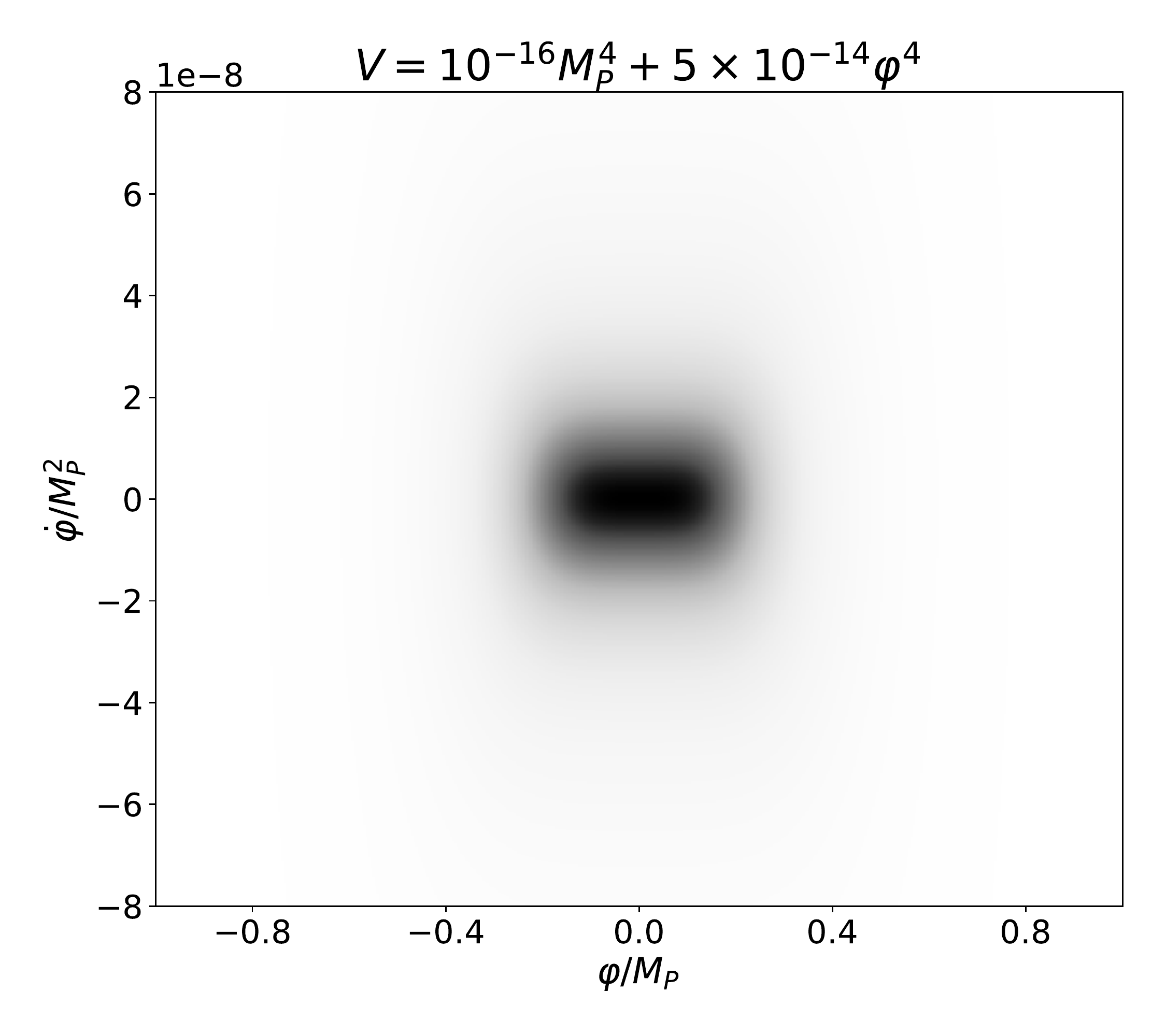}
        \end{subfigure}}
            \caption{Distribution of the values of $\varphi$ and $\dot{\varphi}$ on the phase-space manifold for the potential $V(\varphi)=\lambda\varphi^4+\Lambda$. We have $\lambda=5\times10^{-14}$ for both plots as fixed by the amplitude of scalar perturbations in the CMB, and $\Lambda=10^{-15}M_P^4$ and $\Lambda =10^{-16}M_P^4$ for the left and right plots respectively. Notice how reducing $\Lambda$ causes the distribution to become more tightly clustered around $\varphi=0$. Darker areas correspond to a higher value of $P$ as given in~\eqref{eq:probability}.
            \label{fig:distribution}}
\end{figure*}    
 \end{centering}

We can now calculate the distribution of $\varphi$ and $\beta$ on the phase-space manifold by using the general expression for the distribution given in~\eqref{eq:probability}. This is shown in Figure~\ref{fig:distribution}, where~$\beta$ is expressed in the more physically understandable coordinate~$\dot{\varphi}$ using~\eqref{eq:pdot of beta}. Note that the distribution is peaked close to $\varphi=0$ and tails off very quickly for large $\varphi$. This is to be expected, since the measure goes like $\varphi^{-2}$ for large $\varphi$.

Note also that reducing the cosmological constant sharpens the peak of the distribution.
We argue that this prediction is generic and not specific to our choice of potential. For any potential with a global minimum, as the value of that minimum (i.e. $\Lambda$) tends to zero, the fraction of phase space where the field is found at the minimum tends to one. This can be seen from the form of $P$ in~\eqref{eq:probability}, which diverges as $V(\varphi)\to 0$.

We now wish to calculate the fraction of phase space that allows for $N>60$ e-foldings of inflation to occur, which is required to solve the flatness and horizon problems. As mentioned in the Introduction, regardless of the initial conditions, the Universe will very rapidly evolve into the slow-roll attractor solution. From then on, we can analytically solve the Friedmann equations (\ref{eq:friedmann phi}-\ref{eq:unlifted Hamiltonian constraint}) and obtain the number of e-foldings using the the well-known expression
\begin{equation}\label{eq:e-folds}
N=\int_{\varphi_f}^{\varphi_i}\frac{V(\varphi)}{V'(\varphi)}d\varphi,
\end{equation}
where $\varphi_i$ is the value of $\varphi$ when slow roll begins and $\varphi_f$ is its value when slow roll ends, determined by $\epsilon(\varphi_f)=1$. Since $\varphi_f$ is fixed by the potential, the number of e-foldings is dependent only on $\varphi_i$, and thus for a given potential there is some critical value of the field $\varphi_c$ where ${N(\varphi_i=\varphi_c)=60}$. The Universe must reach the slow-roll attractor with $\varphi_i>\varphi_c$ (or $\varphi_i<-\varphi_c$ for symmetric potentials such as we are considering) if we are to get sufficient inflation. For the quartic potential~\eqref{eq:phi_4}, the critical value is ${\varphi_c=22.09}$ in Planck units. This value is very insensitive to both the cosmological constant $\Lambda$ and the quartic coupling $\lambda$. It is insensitive to $\Lambda$  since inflation occurs at large values of $\varphi$ where the cosmological constant is negligible. Moreover, when $\Lambda$ can be neglected, $\lambda$ cancels in the calculation of $\epsilon$ as well as in~\eqref{eq:e-folds}. Thus, $\varphi_c $ is also insensitive to $\lambda$.

The inflationary attractor is so effective that a universe that starts with $\varphi>\varphi_c$ will reach the attractor with $\varphi_i>\varphi_c$ for all but a very small range of initial value for $\dot{\varphi}$ (or equivalently initial value of $\beta$).\footnote{Note that this statement is only well defined because we have an unambiguous distribution of $\dot{\varphi}$.} Similarly, a universe that starts with $\varphi<\varphi_c$ can only to move to the region $\varphi>\varphi_c$ before slow-roll takes over if $\dot{\varphi}$ is exponentially large (or $\beta$ is exponentially close to $\pi$) so as to overcome the Hubble friction. We therefore make the simplifying assumption that slow roll begins immediately and thus, the fraction of phase space that allows more than 60 e-foldings of inflation is
\begin{equation}
P(N>60)\approx P(\abs{\varphi}>\varphi_c).
\end{equation}

\begin{figure*}
\includegraphics[width=\textwidth]{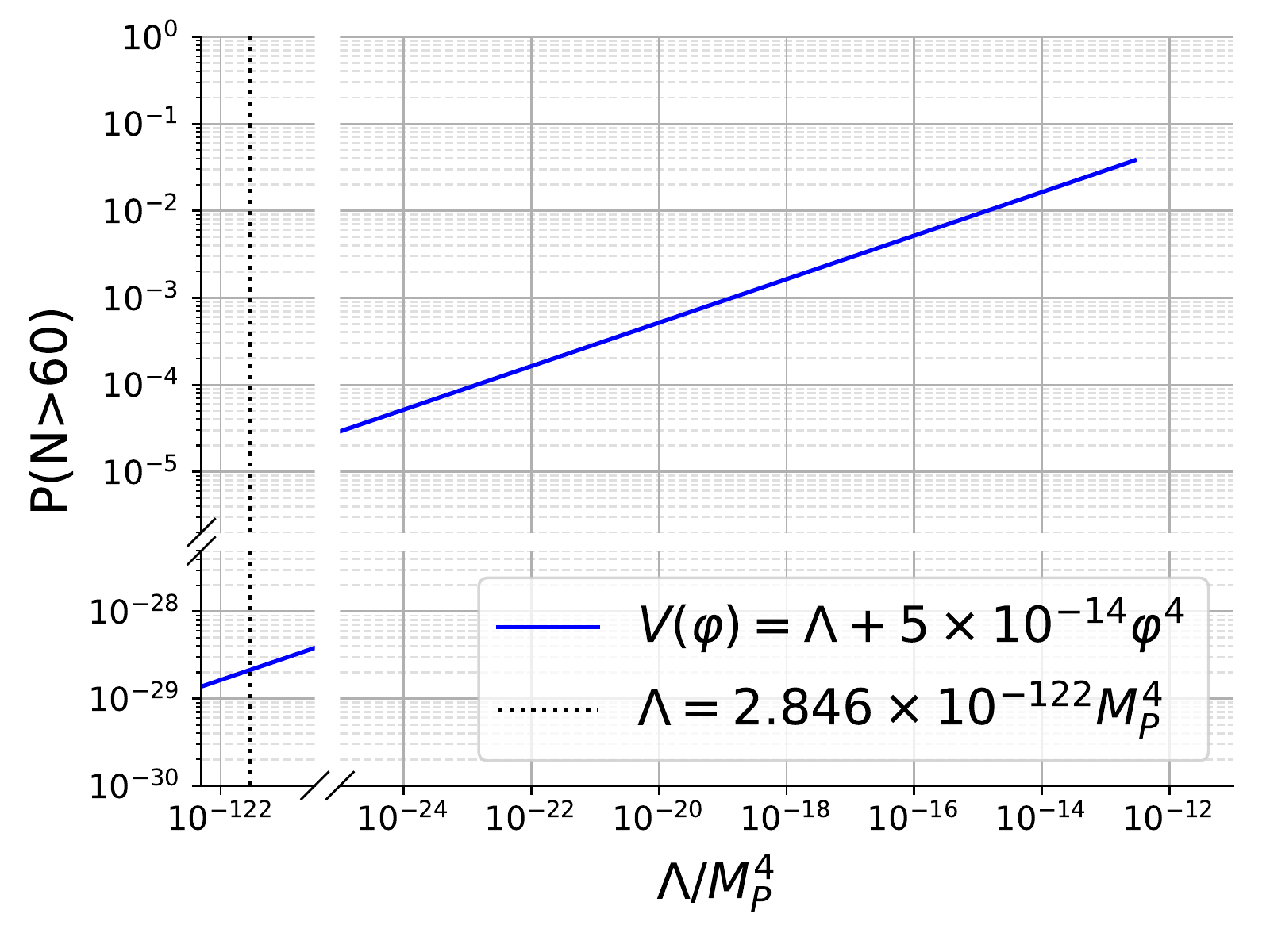}
\caption{Fraction of phase space that allows $N>60$ e-foldings of inflation as a function of the cosmological constant $\Lambda$, with potential $V(\varphi)=\lambda\varphi^4+\Lambda$. The amplitude of scalar perturbations in the CMB has been used to set $\lambda=5\times10^{-14}$. A vertical line is placed at $\Lambda=2.846\times 10^{-122}M_P^4$ corresponding to the observed value of the cosmological constant today. \label{fig:p_infl}}
\end{figure*}

For the potential~\eqref{eq:phi_4}, the fraction of phase space that allows inflation has an analytic expression in terms of hypergeometric functions:
\begin{equation}\label{eq:analytic p_infl}
P(\abs{\varphi}>\varphi_c)=\frac{\sqrt{\pi } \, _2F_1\left(\frac{1}{4},\frac{1}{2};\frac{5}{4};-\frac{\Lambda }{\lambda  \varphi_c }\right)}{4 \varphi_c  \Gamma \left(\frac{5}{4}\right)^2 \sqrt[4]{\frac{\lambda }{\Lambda }}}.
\end{equation}
This is shown in Figure~\ref{fig:p_infl} as a function of the cosmological constant $\Lambda$.

We see that universes with $N>60$ e-foldings of inflation are not generic and hence require fine-tuning of the initial conditions. We observe that the amount of fine tuning required depends on the value of the cosmological constant. If one can construct a model in which the cosmological constant is much larger in the past than it is today, it is possible that the fine tuning may only be percent level for $\lambda\varphi^4$ models. However, for smaller values of the cosmological constant, the fine-tuning becomes much more pronounced. If we assume the cosmological constant has not changed between the early Universe and today, we must use~\eqref{eq:analytic p_infl} with ${\Lambda=2.846\times10^{-122}}$. We then find that the fraction of phase space that allows~60 e-foldings of inflation is ${2.1\times10^{-29}}$.

Is this fine tuning specific to this potential or is it generic? It is impossible to be certain without testing every potential individually. However, it is not difficult to argue that there will be at least some degree of fine tuning required to achieve inflation for most simple potentials. This is because sufficient inflation requires an initial field value that is typically far from the minimum of the potential, whereas the distribution we have derived prefers the initial value to be close to the minimum due to the $V^{-\frac{1}{2}}$ enhancement of the volume element. This property is exacerbated for smaller values of the cosmological constant as argued above.

It is important to note that all the probabilities calculated so far are less than $50\%$, even with the most favourable choice of parameters. This means that, generically, most universes do not inflate as much as usually required to explain the flatness and horizon problems. In light of this result, we are faced with an uncomfortable truth: if we cannot design a potential that avoids this issue, we must work hard to explain why we find ourselves in one of the minority of universes where there is enough inflation. Otherwise, we will be forced to abandon inflation as a solution to the problems of standard cosmology and seek an alternative explanation for the horizon and flatness problems.

\section{Discussion}
\label{sec:discussion}

We have applied the Eisenhart-lift formalism to a theory with gravity and a single scalar field in the minisuperspace approximation. We have used the resulting theory to construct a phase-space manifold, complete with a natural metric whose points represent all possible initial conditions for inflation. The total volume of this manifold is finite when the physically irrelevant directions are integrated over. We have used the natural measure on this manifold to discriminate between finely tuned and generic sets of initial conditions for an inflationary theory without the need for a regulator. We have found that the regions of phase space that allow for $N>60$ e-foldings of inflation make up a tiny fraction of the manifold.

\newpage
We wish to emphasise three key elements of this paper:
\begin{enumerate}
\item This is the first practical application of the Eisenhart lift for field theories.
\item To our knowledge, this is the first time that the Sasaki metric has been applied to the phase-space manifold and the first time~\eqref{eq:phase space metric} appears in the literature.
\item The natural measure which we have constructed on the space of initial conditions does not require regularisation, in contrast to previous treatments in the literature. 
\end{enumerate}

This final point bears further elaboration. Previous authors have placed a natural, finite measure on the set of initial conditions with the help of a regulator. Although their final results are independent of the value of their regulator, they are not independent of the regulation scheme \cite{Hawking:1987bi}. That is because when the total measure of the system is infinite, it is not clear how to use it to define ratios in an unambiguous way. We therefore argue that any measure used to distinguish generic initial conditions from finely tuned ones must be finite in total for these comparisons to be rigorous. The measure presented in~\eqref{eq:full measure} satisfies this requirement. Although it does contain divergences, these are only in physically meaningless directions which arise as a result of redundancies in our description of the initial conditions. With these redundant degrees of freedom integrated out, the distribution presented in~\eqref{eq:probability} is well-defined and finite.


We have been very careful to avoid the use of the term ``probability of inflation'' up until this point, choosing instead to speak of the ``fraction of phase space leading to inflation'', and for good reason. Identifying a measure as a probability measure represents a conceptual step that must be justified. 
We now make this step with the following justification: at very early times, the Universe should be described by more complete theory that must include quantum gravity. At some point, General Relativity becomes a good approximation, and from then on, we are able to track the evolution of the Universe using known physics. It is the value of the fields (and their derivatives) at the end of this quantum gravity regime that sets our initial conditions. 

If we had a theory of quantum gravity, it would presumably be able to tell us what initial conditions we should expect. However, we do not yet have such a theory, and hence we choose to assume as little as possible about the quantum gravity regime. This motivates us to use Laplace's principle of indifference~\cite{laplace1952philosophical}, which states that each possible outcome should be considered equally likely. We can only make use of this principle because the volume of the phase space manifold is finite thanks to our measure. We can thus think of the quantum gravity regime as a \emph{blind creator} ``throwing darts'' at the phase space manifold in order to choose the initial conditions. With this picture in mind, the distribution~\eqref{eq:probability} can then be viewed as a probability distribution. 

We note that the above follows a \emph{frequentist} approach to probability. Our distribution is interpreted as the probability for the Universe to emerge from the quantum gravity regime with a particular configuration. In this viewpoint, the birth of the Universe is simply a probability trial. If we were to perform multiple trials, sufficient inflation would on average occur a number of times proportional to its probability. Indeed, this may have already happened; if the early Universe is made up of several causally disconnected patches, each patch would represent a trial. 

On the other hand, we can approach the notion of probability from a \emph{Bayesian} standpoint. In this interpretation, the probability would parametrise our ignorance of whether inflation occurred in our Universe. In order to calculate the likelihood of inflation, we would use current observations as a condition to update our priors. However, the Bayesian approach has an additional subtlety. Since no observations can distinguish between sets of initial conditions that evolve into each other under time evolution, we must count \emph{trajectories} on the manifold, not points. This may be achieved by defining a counting surface through which each trajectory passes once and only once. One may then use the volume element of this submanifold as a probability density distribution~\cite{Gibbons:1986xk,Gibbons:2006pa}.

It is worth noting that the Bayesian approach introduces a new arbitrary choice, that of which counting surface to use. Although one may expect that the total Liouville measure taken up by a set of trajectories should not change with time due to Liouville's theorem, this actually breaks down when we regularise, even if the regularisation is only in a gauge direction~\cite{Corichi:2013kua}. In fact, the mere existence of an attractor solution tells us that the measure cannot be constant in time.

We have chosen to adopt a frequentist approach,\footnote{Note that we can use a frequentist approach only because the total volume of the phase-space manifold is finite. Previous analyses with an infinite phase space have necessarily adopted a Bayesian approach with a finite trajectory-counting surface.} inspired by an argument by Schiffrin and Wald \cite{Schiffrin:2012zf}.\footnote{See also \cite{ECKHARDT2006439} for further discussion.} We know that the EoMs of the Universe are symmetric under time reversal. Thus, if we use Bayesian probability to \emph{retrodict} the past state of our Universe, we are guaranteed to obtain the same result as when we use it to predict the future.  However, we must remember that our Universe is governed by the second law of thermodynamics. Therefore, while Bayesian probability will correctly predict that the Universe will have higher entropy in the future, it will falsely retrodict that the Universe had higher entropy in the past. In fact, the ``true'' trajectory of the Universe, which had lower entropy in the past, is sure to be assigned a very low Bayesian probability because it is vastly outweighed by trajectories that had high entropy in the past.

Within the frequentist interpretation, we can see that our results show that the probability of sufficient inflation is very small if we assume that quantum gravity leads to a uniform distribution of initial conditions. Therefore, we have to ask that any future theory of quantum gravity leads to a distribution of initial conditions very different from uniform on the phase space if inflation is to continue being a viable theory. In particular, we must insist that the distribution be strongly peaked in the small region of phase space that leads to sufficient inflation. This is a strong constraint that may well prove very challenging to satisfy. 

Inflation may instead be rescued by carefully designing a potential for which the acceptable region is no longer small. A possible example would be a potential with a very large, or even infinite, plateau at $V=V_p$. If the plateau is long enough, it may be able to overcome the~$\sim \sqrt{\Lambda/V_p}$ suppression coming from the determinant of the phase-space metric and take up a significant portion of the manifold. However, if the plateau is truly infinite, then the total volume of the phase-space manifold will no longer be finite. In this case, a new approach would be required to determine if the theory requires finely tuned initial conditions. This is a potentially interesting avenue for future work on the subject.

These challenges, whilst not insurmountable, do place doubts on inflation's ability to solve the problems it was designed for. If inflation is just swapping one set of fine tuning for another, it behooves us to ask if it really does fulfil its goal of resolving the fine-tuning problems of standard cosmology. It may be time to re-evaluate the advantages of inflation and investigate whether there is another theory that might better explain the classic cosmological puzzles of the Hot Big Bang model.

\begin{acknowledgements}
The authors would like to thank Fedor Bezrukov, Jack Holguin, Apostolos Pilaftsis, Chris Shepherd and David Sloan for useful comments and discussions. K.F. is supported by the University of Manchester through the President's Doctoral Scholar Award. S.K. is supported by STFC via the Lancaster-Manchester-Sheffield Consortium for Fundamental Physics,  ST/L000520/1.
\end{acknowledgements}

\begin{sidewaysfigure}
\appendix
\section{Phase Space Metric for Inflation}
\label{sec:app phase space metric}
In coordinates ${\Phi^\alpha=\{a,\varphi,\chi,\dot{a},\dot{\varphi},\dot{\chi}\}}$, the phase-space metric is given by
\begin{equation}
\G_{\alpha\beta}=\left(
\begin{array}{cccccc}
 \frac{9 \dot{\chi }^2}{4 a^5 V}+a \left(\frac{9 \dot{\varphi }^2}{4}-6\right)-\frac{3 \dot{a}^2}{2 a} & \frac{3 \dot{\chi }^2 V'}{4 a^4 V^2}+\frac{3}{2} a \dot{a} \dot{\varphi } & \frac{3 \dot{\chi } \left(a \dot{\varphi } V'+2 \dot{a} V\right)}{2 a^5 V^2} & -3
   \dot{a} & \frac{3 a^2 \dot{\varphi }}{2} & -\frac{3 \dot{\chi }}{2 a^4 V} \\
 \frac{3 \dot{\chi }^2 V'}{4 a^4 V^2}+\frac{3}{2} a \dot{a} \dot{\varphi } & \frac{\dot{\chi }^2 \left(V'\right)^2}{4 a^3 V^3}+a^3 \left(1-\frac{3 \dot{\varphi }^2}{8}\right)+\frac{9 \dot{a}^2 a}{4} & \frac{\dot{\chi } \left(3 a V^2 \dot{\varphi }+2 a \dot{\varphi }
   \left(V'\right)^2+12 \dot{a} V V'\right)}{8 a^4 V^3} & -\frac{3}{2} a^2 \dot{\varphi } & \frac{3 a^2 \dot{a}}{2} & -\frac{\dot{\chi } V'}{2 a^3 V^2} \\
 \frac{3 \dot{\chi } \left(a \dot{\varphi } V'+2 \dot{a} V\right)}{2 a^5 V^2} & \frac{\dot{\chi } \left(3 a V^2 \dot{\varphi }+2 a \dot{\varphi } \left(V'\right)^2+12 \dot{a} V V'\right)}{8 a^4 V^3} & \frac{2 a^6 V \dot{\varphi }^2 \left(V'\right)^2-3 V^2
   \left(\dot{\chi }^2-4 a^5 \dot{a} \dot{\varphi } V'\right)+2 \left(4 a^6+9 \dot{a}^2 a^4\right) V^3+2 \dot{\chi }^2 \left(V'\right)^2}{8 a^9 V^4} & \frac{3 \dot{\chi }}{2 a^4 V} & \frac{\dot{\chi } V'}{2 a^3 V^2} & -\frac{a \dot{\varphi } V'+3 \dot{a} V}{2
   a^4 V^2} \\
 -3 \dot{a} & -\frac{3}{2} a^2 \dot{\varphi } & \frac{3 \dot{\chi }}{2 a^4 V} & -6 a & 0 & 0 \\
 \frac{3 a^2 \dot{\varphi }}{2} & \frac{3 a^2 \dot{a}}{2} & \frac{\dot{\chi } V'}{2 a^3 V^2} & 0 & a^3 & 0 \\
 -\frac{3 \dot{\chi }}{2 a^4 V} & -\frac{\dot{\chi } V'}{2 a^3 V^2} & -\frac{a \dot{\varphi } V'+3 \dot{a} V}{2 a^4 V^2} & 0 & 0 & \frac{1}{a^3 V} \\
\end{array}
\right).\nonumber
\end{equation}
\end{sidewaysfigure}

\end{document}